# Near-IR Spectrograph for VSI (VLTI Spectro-Imager) dispersing the light from an integrated optics beam-combiner

Dario Lorenzetti*[a], Gianluca Li Causi[a], Roberto Speziali[a], Fabrizio Vitali[a], Davide Loreggia[b], Carlo Baffa[c], Pierre Y. Kern[d], Laurent Jocou[d], Fabien Malbet[d], Patrick Rabou[d]

[a]INAF-Osservatorio Astronomico di Roma - Via Frascati,33 – 00040 Monte Porzio (Italy);
[b]INAF – Osservatorio Astronomico di Torino, via Osservatorio, 20 – 10025 Pino Torinese (Italy)
[c]INAF – Osservatorio Astrofisico di Arcetri - Largo E.Fermi, 5 – 50125 Firenze (Italy)
[d]Université J.Fourier, CNRA, Lab. d'Astrophysique de Grenoble, UMR 5571, BP 53, F-38041 Grenoble cedex 9 (France)

## ABSTRACT

We present the optical and cryo-mechanical solutions for the Spectrograph of VSI (VLTI Spectro-Imager), the second generation near-infrared (J, H and K bands) interferometric instrument for the VLTI. The peculiarity of this spectrograph is represented by the Integrated Optics (IO) beam-combiner, a small and delicate component which is located inside the cryostat and makes VSI capable to coherently combine 4, 6 or even 8 telescopes. The optics have been specifically designed to match the IO combiner output with the IR detector still preserving the needed spatial and spectral sampling, as well as the required fringe spacing. A compact device that allows us to interchange spectral resolutions (from R=200 to R=12000), is also presented.

## 1. INTRODUCTION

Aiming to investigate astrophysical phenomena that occur at the smallest spatial scales (milliarcsec), the observer has to combine, in principle, a very high spatial resolution with enough spectral resolution to resolve the line profile. This latter requirement could be partially relaxed but still preserving the capability of identifying spectral features, that represent invaluable tracers of any individual environment. In other words, the spectroscopical modality allows us to identify a particular physical region whose size is determined by means of a high spatial resolution modality. This result can be effectively obtained by dispersing the light once it has been coherently combined from several large aperture telescopes. This is the working principle of an interferometric spectro-imager like VSI. An exhaustive discussion about the application of this general concept along with many details about the VSI system can be found elsewhere in this issue (Malbet et al. 2008 [1]; Jocou et al. 2008 [2]).

The present paper deals with the design of the VSI IR spectrograph, namely the cold (77 K) opto-mechanical components aimed to disperse the radiation onto the IR array. More than the technical concept is provided, going into specific details about glasses, tolerances and manufacturing feasibility. A particular care has been payed to make the system as much simple and inexpensive as possible. The presented opto-mechanical design is able to fulfill the scientific requirements and the technical items are all optimized for exploiting, wherever is possible, standard and tested solutions. In Sect.2 the input conditions will be shortly summarized, while in the following Sect. 3, 4 and 5 the optical, cryogenic and mechanical aspects are discussed, respectively. Finally, in Sect.6 the mechanical assembly for positioning and aligning the IR detector is described.

*dloren@mporzio.astro.it; phone +39 06 94286435; fax +39 06 9447243.

## 2. INPUT CONDITIONS

We summarize here the relevant input specifications on which the present design is based:

- Wavelength range: 1.08 - 2.38 μm (J, H and K bands). J:1.08-1.35 μm, H: 1.5-1.8 μm, K: 1.95-2.38 μm.
- Interchangeable modes at different spectral resolution: Low res. (LR) variable $R \sim 30$ to 200 (the need of a variable modality has to be still confirmed); Medium res. (MR) $R \sim 2000$; High res. (HR) $R \sim 12000$. A mode at intermediate resolution (IR) $R \sim 5000$ should be considered only if it does not imply a major impact on the instrument design.
- Three dedicated (JHK) multi-axial beam combiners kept at a working temperature of -40 °C.
- Magnification: 5.88 in J, 7.4 in H, and 9.0 in K band.
- Image quality: point spread function in one pixel.
- A cold plate is required in front of the detector for carefully evaluating the detector dark current.
- A cold plate at the slit level for evaluating the system dark.
- Flat field at pupil location.
- Power dissipated in the interferometric laboratory: ESO specifies 10W of convected heat load and 20W for the total convected cooling load from each instrument inside the interferometric laboratory.
- Outer surface temperature: all equipment located in the laboratory shall not have an outer surface temperature different by more than +0.5 / -1 °C from the ambient air temperature.
- Avoidance of any cryo-coolers.

## 3. OPTICAL DESIGN

### 3.1 Overall layout

The spectrograph layout has been designed to be compatible with both the 4 way and the 6 way beam combiners. An overall sketch of this layout is given in Figure 1. The light enters the system at one of the Integrated Optics (IO) combiners kept at -40°C and fed by means of fibers (depicted in the left upper corner of Figure 1). The optical design considers three collimators each optimized for a specific band (J, H and K). Such configuration, which allows to keep both the beam combiners with their feeding fibers and the collimators themselves in fixed positions, needs of an additional movable flat mirror for selecting the working optical train. The different collimators are designed to form the same slit image on the detector. Moreover, this solution permits to avoid a movable filter wheel inside the cryostat, since any individual filter for selecting the adequate bandpass is mounted in the dedicated collimator assembly.

### 3.2 Integrated optics beam-combiner

The VSI technological asset is represented by the chips of integrated optics that provide the beam combination; a more detailed discussion on this aspect is given in Jocou et al. 2008 [2] . This approach allows an optical device to be embedded in, or deposited upon, a substrate by using photo-lythography techniques. In this way, complex optical circuits can be implemented within small chips (few squared centimeters) and many studies and realizations stem from the telecommunication industry which has to cope with an increasing demand in terms of physical sizes and bandwidth. An optical combiner is constituted by three elements (i) connectors to plug the fiber bundle; (ii) fibers to carry the incoming light to the chip; (iii) integrated optics (IO) chip where the interferometric pattern originates. Using an IO circuit to combine an array of telescopes gives important advantages concerning compactness, stability, self-alignment, flexibility in swithching beam combination configuration. A sketch of this device is displayed in Figure 2.

A single-mode waveguide with the proper dimension allows to extract the coherent part of the signal, thus accomplishing the so-called *modal (or spatial) filtering*. Wavefront rugosities will be projected on the waveguide modes, but only the energy carried by the fundamental mode will be propagated all the way to the combiner. Simulations show that higher order modes cannot propagate and are evacuated rapidly. The resulting wavefront has a gaussian intensity with an excellent flatness. Such a flatness combined with a proper monitoring of photometric fluctuations ensures a very accurate interferometric calibration which has given, so far, the best visibility accuracies ever measured [2].

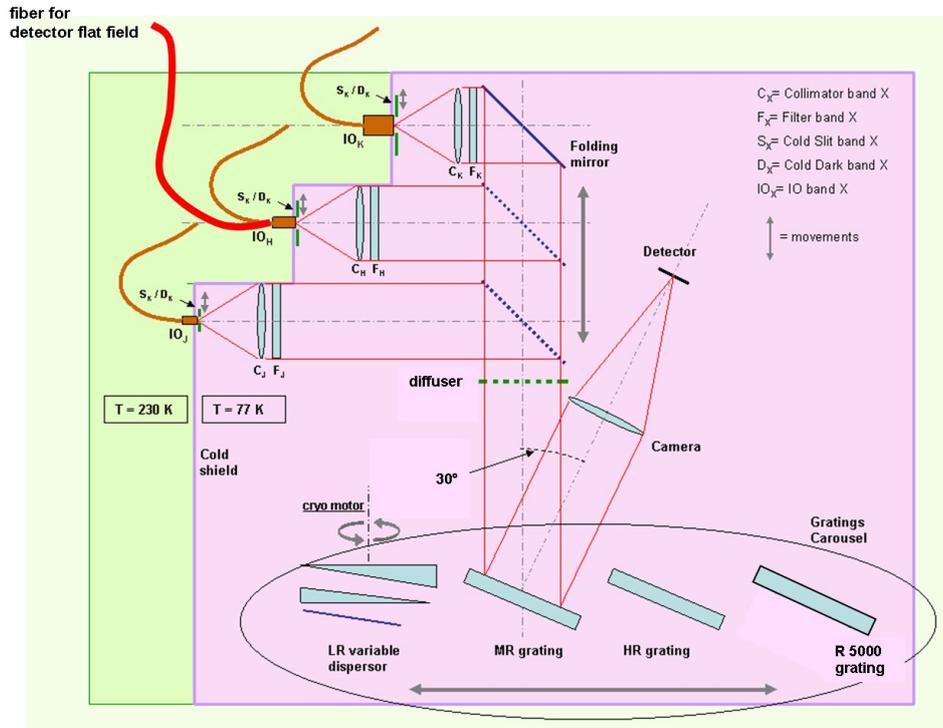

Fig. 1. Sketch of the overall spectrograph layout. Each beam combiner has its own optimized collimator.

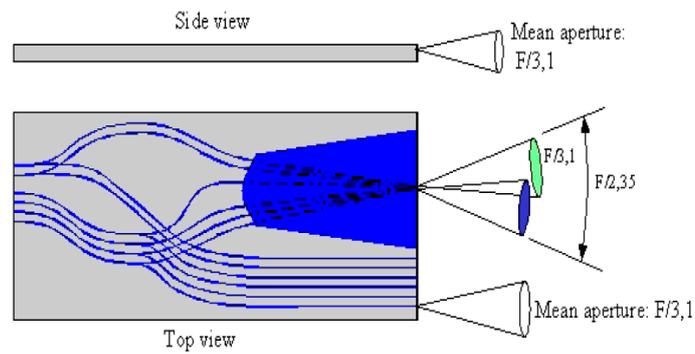

Fig. 2.  Beam combiner output features.

### 3.3 Pupil analysis

The multi-axial IO system can be considered, from the optical point of view, as composed by two sets of point-like sources, i.e. the photometric outputs and the multi-axial area, producing different types of beams (see Figure 2). With reference to Figure 3, calling X the direction of the IO plane and Y the orthogonal one, we have:

- the photometric outputs: six point-like sources located at the edge of the IO, coming out with telecentric circular gaussian F/3.1 beams at $1/e^2$ intensity radius;
- the multi-axial output: one X-linear source located at the edge of the IO, each point of which comes out with a complex beam shape, composed by six non-telecentric Y-linear gaussian F/3.1 beams at $1/e^2$ intensity width, each one coming from the virtual image of the corresponding multi-axial input.

The collimator will produce two different pupils from such complex inputs, as depicted in Figure 3:

- a photometric gaussian circular pupil, located at focal distance because of the telecentricity of the photometric outputs;

- a multi-axial striped pupil, located at the conjugate plane of the virtual image of the multi-axial input, i.e. beyond its focal distance. Only the first of these pupils is present in a pairwise 4 beams IO combiner.

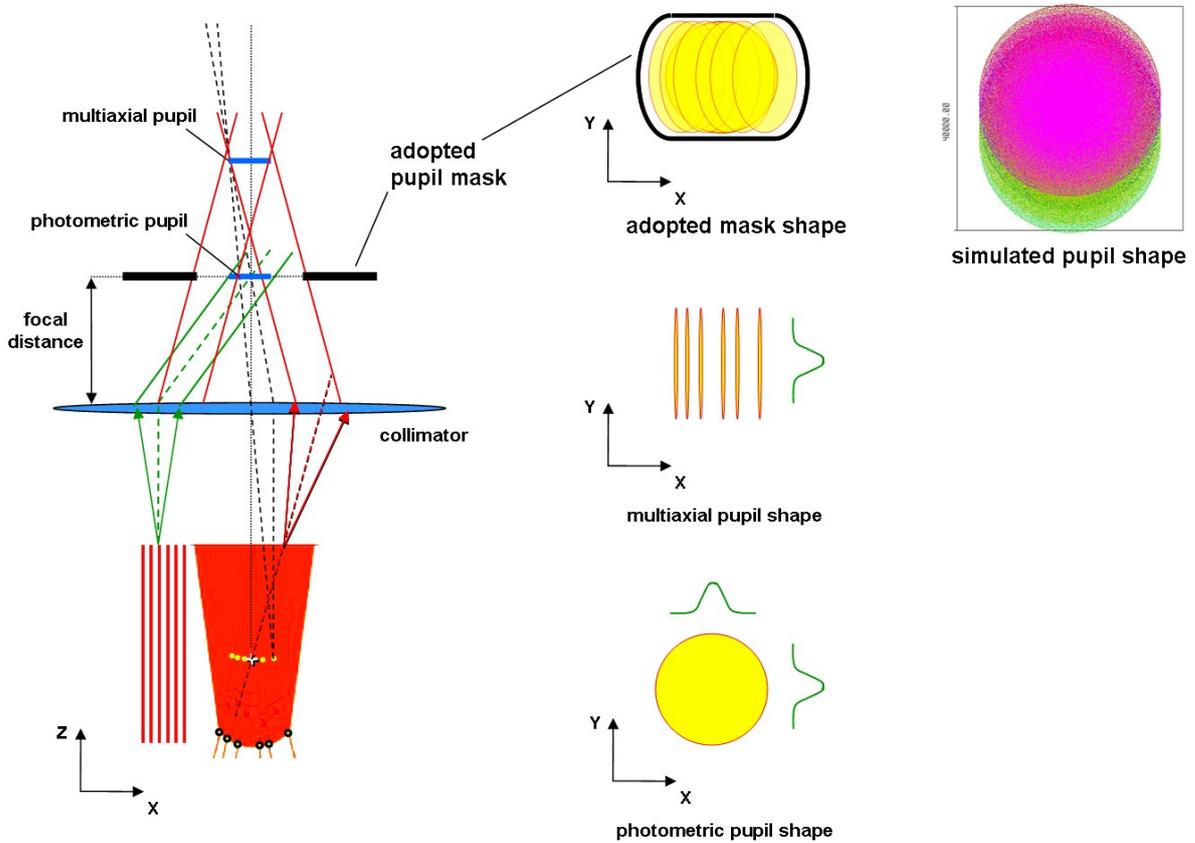

Fig. 3. Shape and location of the photometric and multi-axial pupils (left); shape of the adopted pupil mask to be placed at the collimator focal distance (center); ZEMAX simulation of the pupil shape (right).

To have an end-to-end simulation of the spectrograph optical system, we have realized a ZEMAX simulation of the 6T IO output beams (see Figure 4), based on the IO dimension. In Figure 5 a 3D view of the output beams clearly shows the different cones of the photometric and multi-axial beams. ZEMAX cannot easily simulate the multi-axial output due to the lack of a built-in mono-mode waveguides models, so that we have assumed the beam parameters computed by a custom software developed in LAOG [2]. Hence, we do not simulate the IO combination itself, but only the output beams, not taking into account any interference effect, but only the incoherent light.

The narrow stripes of the multi-axial pupil allow us, in principle, to put an effective and properly shaped pupil mask, but it would block the circular photometric beams. Hence, unless to use two different optics for the two inputs, it is impossible to adopt such an arrangement. After that some solutions have been comparatively evaluated (Lorenzetti & Li Causi 2007 [3]), we have finally adopted the photometric pupil baffle, with a slightly elongated shape which mainly acts as a stray light baffle. Indeed, an accurate analysis of both the thermal emission and ghosts coming from the IO itself has shown that they are negligible, due to the low emissivity of the IO system at the working temperature of $T_0$= 230 K and to the similarity between the refraction indexes of the IO core and cladding ($n_{core}$= 1.454, $n_{cladding}$ = 1.444 at $\lambda$ =1.5 μm).

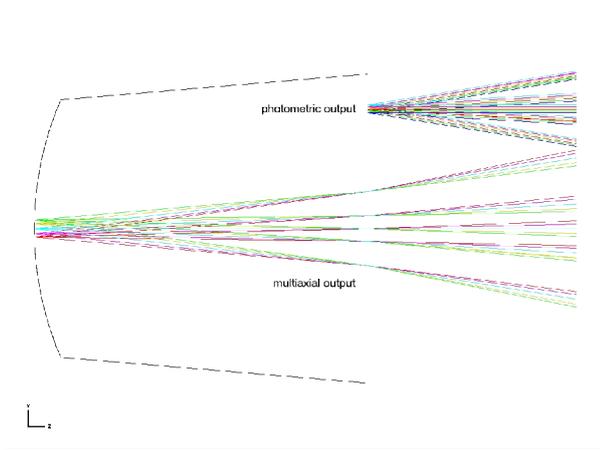 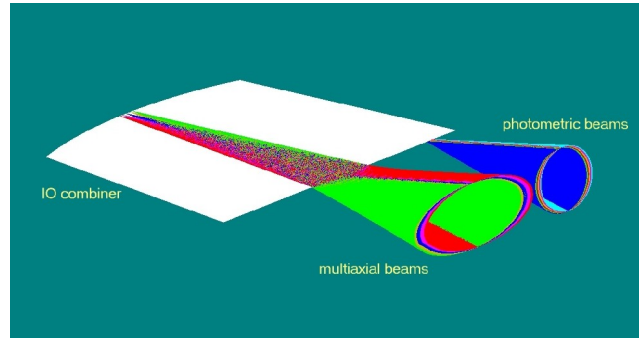

Fig. 4. ZEMAX simulation of the light cones exiting out of the 6T combiner (top view).

Fig. 5. 3D representation of the two kinds of output beams.

### 3.4 Spectrograph optics

The optical design of the spectrograph stems from the specification listed in Sect.2 and it is depicted Figure 6. The concept is based on a set of three collimators each optimized for one spectral band i.e. one beam combiner. The camera optics is then common to the three collimators. In order to match the spectral sampling in the three bands, the magnification should be adapted to $J = 9$, $H = 7.4$, and $K = 5.8$. The pupil diameter is 40 mm to reach the required spectral resolution. The optics are furthermore optimized to collect the extreme F/2.35 aperture of the multi-axial output. Each collimator is composed by a set of two doublets, while the camera uses only one doublet. The features of the optical surfaces of this layout are described Table 1. As shown in Figure 7, the optical quality matches the requirement to keep the PSF in one 18 μm pixel of the HAWAII-II detector.

The view shown in Figure 6 shows the whole optical system including a grating, where an angle of 30° is adopted between the collimator and the camera lenses. The overall dimension of this layout is 1000×600 mm. Note that the optical path could be folded a second time to decrease the overall size. Such a layout will be eventually modified in order to fit the mechanical implementation by taking into account:

- integration of the layout in a dewar smaller than 1×1 m;
- an adequate location of the pupil for the three IO beam combiners.

Tab. 1. Parameters of the optical surfaces.

| Surface | Thick. (mm) | Radius (mm) | Ø (mm) | Glass |
|---|---|---|---|---|
| 1 | 11.98 | - | - | - |
| 2 | 9.39 | -15.82 | 11 | s-tih11 |
| 3 | 0.5 | 295.30 | 15 | air |
| 4 | 16.09 | -177.36 | 18 | BaF$_2$ |
| 5 | 74.45 | 1775.05 | 18 | air |
| 6 | 8 | -336.94 | 20 | s-tfm16 |
| 7 | 10 | -77.65 | 20 | air |
| 8 | 16.86 | -85.25 | 24 | CaF$_2$ |
| 9 | 105 | 327.19 | 24 | air |
| 10 | 0 | 327.19 | - | air |
| 11 | 400 | 327.19 | 42 | air |
| 12 | 11.65 | 327.19 | 42 | CaF$_2$ |
| 13 | 8 | 327.19 | 42 | s-tfm16 |
| 14 | 909.91 | 327.19 | 42 | air |
| 15 | Image | - | - | - |

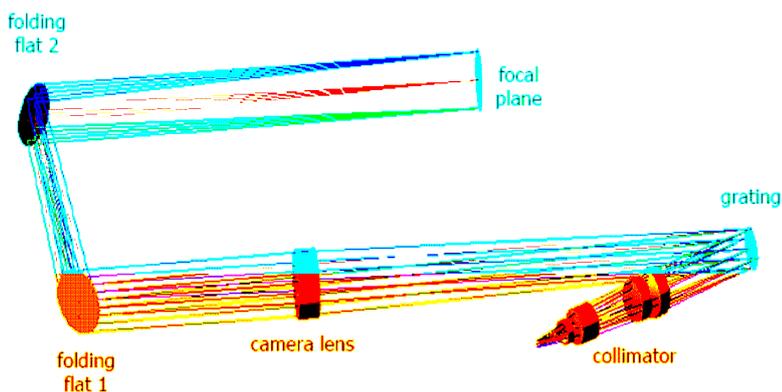
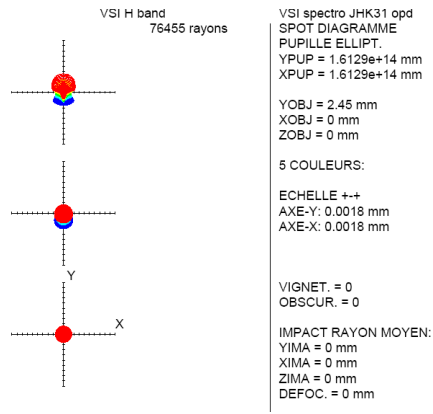

Fig. 6. VSI Spectrograph optical layout.

Fig. 7. Spot diagram (cross size is 36μm).

### 3.5 Calibration devices

***Detector Cold Dark*** - The cold plate will be made by means of a metallic screen at the $LN_2$ temperature, mounted on a simple two-positions mechanical stage between the detector and the last lens of the camera. This is used for detector calibration purposes, such as bad pixel map and flat field map computation.

***Detector Flat Field -*** The flat screen is a transmission diffuser close to the pupil stop position (see Fig.1). A large diameter (Ø = 350μm) multi-mode calcogenide fiber is placed just above the H-band integrated optics, but physically separated from it, in order to use the H-band collimator and filter to send the light to the diffuser. The fiber will be fed by an external continuum lamp placed far away from the instrument or in the source box of the CAT. The diffuser is aimed at producing a Lambertian illumination pattern onto the detector. Its best location is the entrance pupil of the camera, which, in the current optical design, is placed at the same position of the exit pupil of the collimator, i.e. on the pupil baffle. The light from the screen is re-directed towards the camera by means of the medium resolution grating (see below, Sect.4.2), working at its zero order.

***Cold masks*** - Two cold masks are positioned along the spectrograph optical path. The first is an IO chip slit mask. Indeed, the IO chip being cooled down at 230 K can generate thermal emission inside the spectrograph. A 77 K fixed mask is close to, but not in contact with the IO output to limit the emission of the IO chip (see Figure 8). Such a cold mask aims at keeping the thermal contribution (mainly in the K band) significantly lower than the contributions from detector dark current and readout noise. The second (also depicted in Figure 8) is a movable cold plate placed at the same position (namely at the IO pseudo-slit level) and aimed to properly evaluate the dark current upon request.

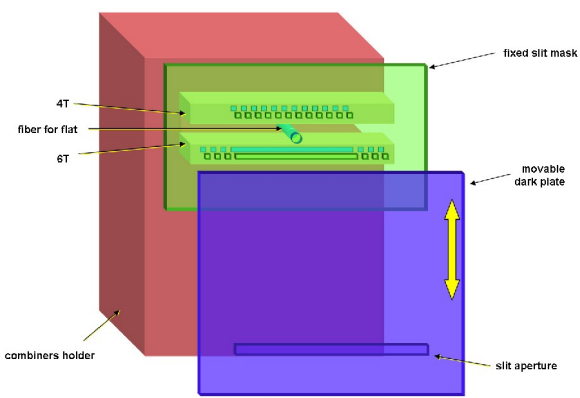

Fig. 8. Cold plate device for both slit mask and dark.

# 4. GRATINGS and FILTERS

The spectral resolutions required by the high level specifications are: variable from R=30 to 200, R=2000, R=12000 and an optional mode at R=5000.

## 4.1 Low resolution disperser

Two concepts have been studied for the realization of a Low Resolution Disperser (LRD) able to provide either a resolution of R = 200 or a very low resolution adapted for faint objects observation. The variable resolution is obtained by combining either a couple of gratings or two prisms doublets in a counter-rotating mount, which allows us to add or subtract their individual dispersions depending on the relative rotation. These concepts are better described below:

• A couple of gratings: a transmission and a reflection grating are combined in a counter-rotating mount, mounted parallel to each other and turning around the same axis (left part of Figure 9). The light coming from the collimator is dispersed two times by the transmission grating and once by the reflective one, the final dispersion being the vector summation of each single dispersion.

• Two counter-rotating prisms: each one being composed by two prisms made of different glasses (Fused Silica and $CaF_2$) with nearly similar refractive indexes but different dispersion (right part of Figure 9). A flat mirror is placed after the prisms to fold the light towards the camera. The final choice will depend on the cryogenic optimization of the optical design: the gratings solution is much more tolerant to deviations from parallel collimation and offers a better optical quality, but has a lower efficiency, while the prisms solution requires a perfect collimation and offers a higher efficiency.

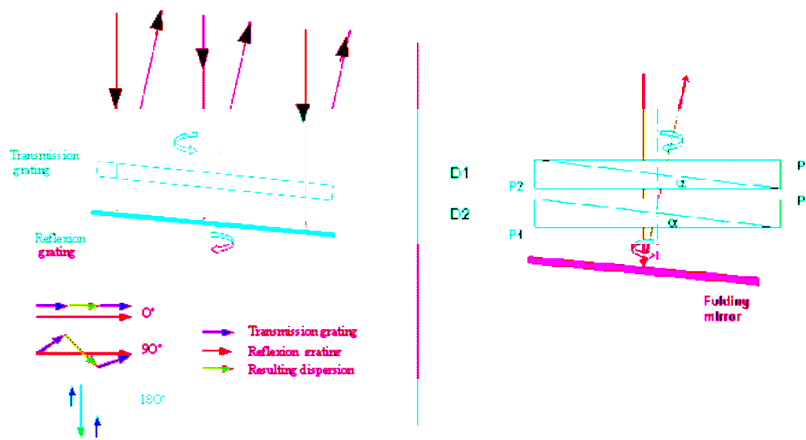

Fig. 9. Concept of the low resolution dispersers.

## 4.2 Medium and High resolution

All the required spectral resolutions are obtained through tunable gratings, whose tunability is provided by the grating turret itself (see Figures 10 and 11). Due to the gaussian pupil, we have an Effective Pupil Diameter $D_{eff} = 0.6 \times D_P / \cos \alpha$ (where $D_P$ is the geometrical pupil diameter and $\alpha$ is the incidence angle); this occurrence causes a lower resolution power of the gratings, without decreasing the angular dispersion. Since the current configuration envisages one single grating for each resolution, then a given band could suffer from a quite poor blaze function. By using standard gratings, their low cost could allow us to adopt two gratings for each resolution, in order to optimize the blaze function: a possible solution could be (as in the medium resolution case as an example) to get one grating optimized for the J band and a second for the H and K bands; this configuration allows 0.75 as minimum value for the blaze function, instead of smaller values imposed by the single grating approach. Obviously, this has an impact on the overall size of the grating turret, but the benefit in terms of blaze function (e.g. efficiency) could be worth to be taken into consideration. This solution is indeed more convenient in term of costs, since a custom grating (even less efficient than the two standard gratings) costs from 25 to 50 times more than a standard one. In the following a set of three gratings is provided, considering the parameters given in Table 2.

Tab. 2: Relevant parameters for the grating selection.

| Array side (pxl) | 2048 |
|---|---|
| Pixel size (μm) | 18.0 |
| Array side (mm) | 36.864 |
| ELF (mm) | 917 |
| Off- angle (deg) | 30 |

where: the Effective Focal Length (EFL) is the camera focal length, Off-angle is the fixed separation angle between the collimator and the camera. In the following calculation we consider standard grating from the Newport-Richardson Catalog.

*Medium Resolution  R ~ 2000*

We adopt J band at second order and the H and K bands at first order. The grating should be tuned to properly center the desired band on the FPA, together with the selection of the relative collimator and filter. In Table 3 the main parameters of the selected grating RGL 53-*-750R  are given.

*Intermediate Resolution R ~ 5000*

With the chosen grating, we have the J, H and K bands at first order, with resolutions of the order 4000-5600. The grating should be tuned to properly center the desired portion of the band onto the IR array: the J band enters entirely in the array, whereas the H and K bands can be covered with two scans each. Both H and K bands have nominal resolution higher than the theoretical resolution. In Table 3 the main parameters of the selected grating RGL 53-*-630R are given.

*High Resolution R ~  12000*

With this grating, we have the J, H and K bands at first order, with resolutions of the order 11000-17000. The grating angle should be tuned to properly center the desired portion of the band onto the IR array: the J, H and K band can be covered with four, two, four scans respectively. Both J and K bands have nominal resolution always higher than the required resolution. In Table 3 the main parameters of the selected grating (RGL 53-*-231R) are given.

Tab.3: Parameters of the gratings selected to have the requested resolutions.

| Parameter | *R ~ 2000* | *R ~ 5000* | *R ~ 12000* |
|---|---|---|---|
| Grooves per mm | 75 | 200 | 497 |
| Blaze angle (deg) | 4.3 | 10 | 34 |
| $\lambda_{blaze}$ (μm) | 2.00 | 1.7 | 2.25 |

### 4.3  Filters

The collimators are coupled with the proper J, H and K broad band filters. In recent years, the need to lower the costs of the standard filters led the international astronomical community to join together to place a single multi-filter order to the producer companies. The previous runs led to filter costs of about 2000 US$ each (2 inches), whereas a custom run can costs from 7000 to 10000 US$. If some commercially available filter will have a bandpass too different from the VSI specification, we should envisage a custom run for that.

# 5. GLOBAL EFFICIENCY

Finally, we can estimate the throughput expected for the spectrograph. The computation is done from the first collimator surface to the last camera surface. The transmission coefficients of the glasses have been considered along with the classical attenuation law within any individual glass. The resulting efficiency values in any spectral band are summarized in Table 4 by accounting for any transmitting or reflecting surface.

Tab.4: Contributions of the spectrograph components to the total throughput.

| Glass | Thick. (mm) | T(%) 1.2μm | T(%) 1.6μm | T(%) 2.3μm | T(%) 2.4μm |
|---|---|---|---|---|---|
| s-tih11 | 9.4 | 0.99 | 0.99 | 0.98 | 0.94 |
| $BaF_2$ | 16 | 0.98 | 0.98 | 0.98 | 0.98 |
| s-ftm 16 | 16 | 0.99 | 0.99 | 0.98 | 0.92 |
| $CaF_2$ | 27 | 0.98 | 0.98 | 0.98 | 0.98 |
| Grating (MR/UR) – at peak | | 0.80 | 0.80 | 0.80 | 0.80 |
| Coating (12 surfaces) | | 0.78 | 0.78 | 0.78 | 0.78 |
| Order Sorter Filter | | 0.85 | 0.87 | 0.90 | 0.90 |
| **Total** | | **0.73** | **0.73** | **0.72** | **0.65** |

# 6. CRYO-MECHANICAL DESIGN

The VSI optical layout is mounted on an aluminum optical bench that is also the back-plate of the nitrogen vessel of the cryostat for having a uniform cooling of the optical components. A 3D design of all the components has been realized and depicted in Figure 10. The positions of the motorized axes have been optimized to have both the internal axes and the external motors in the most convenient arrangement. With reference to Figure 10, the main mechanical components of the VSI spectrograph are those listed below, but we will discuss in the following only the rotating stage (Sect.6.1).

- Three supports of IO with cold slit and dark screen
- Three fixed collimators mounts
- One linear stage for band selection
- One linear stage for flat fielding
- One rotating stage housing the dispersing devices
- One fixed camera mount

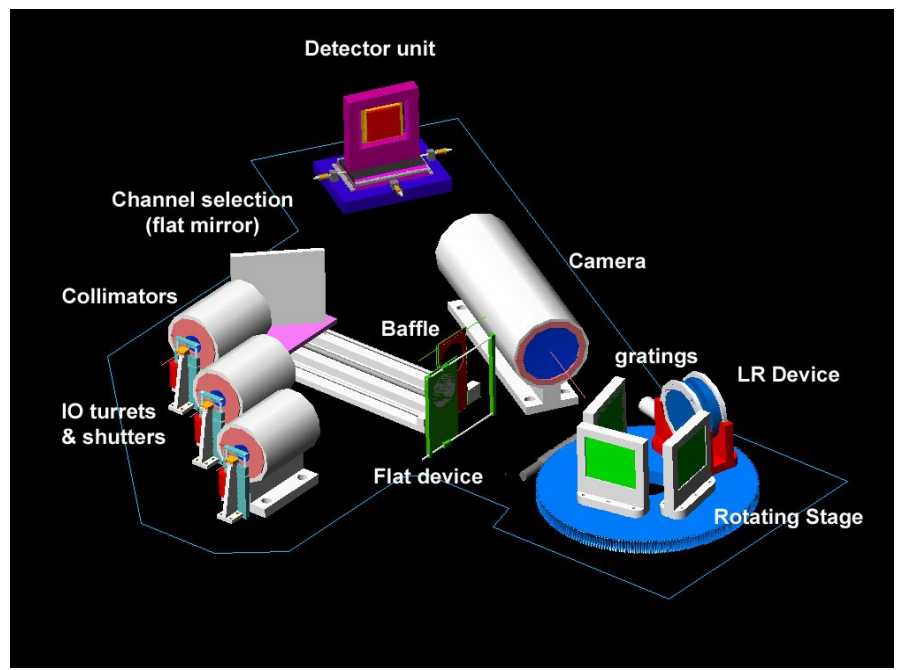

Fig. 10. 3D overall layout of the mechanical assembly of the spectrograph.

### 6.1 Rotating stage

Such unit is a rotating plate supporting all the three gratings at medium, intermediate and high resolution along with the low resolution device (LR). To achieve the positioning repeatability of 5 pixels (i.e 37 arcsec with the current camera) and the required stability of 1/5 of SRE, i.e. 0.4 pixels (about 3 arcsec), we envisage a device driven by a warm motor and able to strongly reduce the backlash. As an alternative to the custom solution based on worm wheels gears, a XYZ cryogenic stage can be evaluated in the next phase. In Figure 11 the 2D design of the turret with all the dispersive devices on its top is given. The gratings are rigidly mounted on the turret and properly oriented to match the required angle between the beam coming from the collimator and the one going toward the camera. In particular, to get the full spectra on the array with the high resolution (R ~12000) grating (HR), a scanning sequence of 4(J)+2(H)+4(K)=10 positions is required (see 4.2). Figure 9 shows how by rotating that grating of an angle = 14° around the turret axis, we are able to cover the entire range; the two limiting positions are indicated in red and blue, respectively. The same rotation also allows for a fine tuning needed to center the H and K bands onto the array in the intermediate resolution (R ~5000) mode. In other words a rotation around the central axis of the turret, driven by a warm motor, allows also for both an adequate scanning motion and for conveniently centering the dispersed bands, thus saving two additional cryogenic motors. The LR device aims to provide a variable dispersion. It is composed by two counter-rotating prisms (or gratings) driven by a cryo-motor and a system of three bevel gears. The maximum rotation angle is 90° (180° difference) and the required precision in about 1°. As the other movable devices, also the gratings and prisms are thermally connected to the cold bench with copper straps.

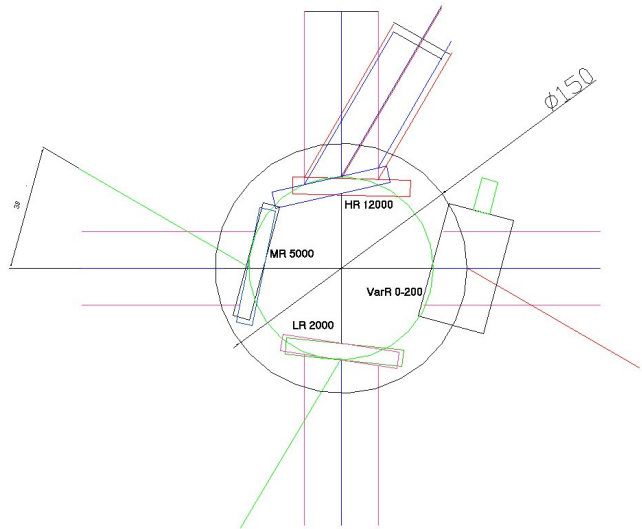

Fig. 11. 2-D sketch of the rotating turret.

### 6.2 Motors and controls

The following motors are required:

- 3 cryo-motors for selecting slit or dark in front of each IO
- 1 room temperature motor for positioning the flat mirror
- 1 room temperature motor for the grating turret (usable for HR scanning, as well)
- 1 room temperature motor for inserting the screen for flat-fielding
- 1 cryogenic motor for the counteracting prisms dedicated to low resolution mode
- 1 room temperature motor for positioning a dark screen in front of the IR array

This means a total of 4 motors at room temperature and 4 cryo-motors. In Table 5 the functions to be motorized are summarized along with the indication of relevant parameters, such as needed travel, number of positions and mechanical resolution. Two types of motors and relative controls used by ESO are under evaluation: the 5-phase Bergher-Lahr motor (used in AMBER) that can be modified to be available as cryogenics, or the more expensive and larger 2-phase cryo-motors from Phytron. A choice between this two models will be done upon the type of use (warm or cold), space, resolution and needed torque.

Tab. 5: Overview of the moving functions.

| Movement | Number | Type | Travel | Position | Resolution |
|---|---|---|---|---|---|
| Slit aperture (or dark) - **cryo** | 3 | linear | 10 mm | 3 | 0.5 mm |
| Flat mirror | 1 | linear | 300 mm | 3 | 0.5 mm |
| Grating turret | 1 | rotating | ± 180° | Cont. | ~ 30 arcsec |
| Counteracting prisms - **cryo** | 1 | linear | ± 90° | Cont. | 1° |
| Illuminated flat | 1 | linear | 50 mm | 2 | 0.5-1 mm |
| Detector dark | 1 | linear | 40 mm | 2 | 0.5-1 mm |

### 6.3 Cryogenic system

The cryostat has not been designed so far, but it will be a quite conventional $LN_2$ cryostat with a foot print of about 1m x 1m x 0.5m. The optical bench that has to be cooled down is actually one side of a vessel ensuring a good thermal contact with the cryogenic liquid. The whole optical bench is enclosed in a radiation shield in tight contact with the cold bench itself. Concerning the pumping process, since the estimated free room inside the cryostat should be comparable with that of AMBER, we can adopt the same evacuation system already in operation, with pressure gauges and temperature controls compatible with ESO requirements.

We make a rough estimate of the thermal balance [4]. We have four main thermal inputs: radiation from the warm enclosure, conduction by the residual gas, conduction through electrical cables and through the bench support. We can estimate an inner radiating surface (vessel + cold shield) of about 17000 $cm^2$ while the outer (warm) surface is of the order of 40000 $cm^2$; by assuming a conservative emissivity values of 0.15 for the 300K surface and 0.12 for the cold one, we can estimate a radiative input of about 60W. If we adopt a multi-layer (>10) Mylar cover mounted over the cold shield we can reach an effective input of the order of 1-2 $W/m^2$, reducing the thermal input by a factor of 7-10.

In the regime of very low pressure (if the particle mean free path is larger than the mechanical gap between inside and outside, in our case under $10^{-5}$ mbar), the thermal conduction through the residual gas is linear with the pressure, and amounts to something of the order of 1 $W/m^2$ · Pressure / $10^{-5}$ mbar. This can led to 4 W of thermal load. The heat flow through the support system can be estimate borrowing the FEA computation for this part from a similar instrument and scaling down: the total is about 5 W. The conduction through the electrical wire can only be guessed, as the cabling is still not defined. Scaling down from a similar instrument, and assuming we can have a mean cable length of about 1m, we have a thermal input of about 5 W. The total thermal load can be estimated to vary from 75 to 20W. Adopting the worst figure, a holding time of 24 hours can be achieved with a consumption of 40 litres of $LN_2$.

## 7. DETECTOR INTERFACE

VSI community has discussed whether or not the double-dewar design, a concept originally adopted for AMBER (one for the spectrograph opto-mechanics and one for the IR array), could still represent a useful alternative to the more common single-dewar solution. This latter presents few but substantial advantages in term of:

- optical design: no optical extension is required; a lesser number of optical elements is needed and the camera lens may be located near the detector
- optical alignment
- mechanical stability and compactness
- management of a lesser amount of interface details have to be agreed
- baffling and shielding the array from stray-light
- reduction of costs

Moreover, following the single-dewar approach, the number of thermal cycles (that could affect the increasing number of bad pixels) will be significantly reduced and the detector can remain in a uncontaminated environment until the final integration. The IR-array will be inserted in the same dewar of the spectrograph that is mechanically developed and cabled in order to accept it. This solution offers the advantage to carefully test the thermal link of the array in its final location, but a separate (may be commercial) cryostat has to be set up for testing the array prior to the final assembly. The preliminary operations of the spectrograph alignment will be done at room temperature by using an optical CCD as a detector and by taking into account the adequate ray tracing for the provisional detector and ambient temperature. Therefore the spectrograph does not need the IR array since the preliminary lab operations, but only during the final integration. The optical tolerances are given in Table 6: the positioning tolerances for the array are referred to a RMS decrease of the spot diagram of 10%.

Tab. 6: Optical tolerances relevant for the detector mounting.

| Movement | Requirement |
|---|---|
| Tilt (on x and y) | 0.3° |
| Tilt (around opt.axis z) | 30 arcsec |
| Piston | +0.1mm; -0.3mm |

Given these tolerance values, mounting of the detector appears a quite straightforward task; indeed, the required tolerances are quite relaxed and can be then satisfied just with the a precision machining, provided a careful thermal study has been done. The de-centering can be easily measured and then adjusted with the camera opened and warm. Therefore, no movable mounting seems to be needed, but only a fixed position for the array. It is worthwhile to consider a warm movement to allow adjustment for de-centering and, possibly, de-focusing. The exact working temperature of the array will be determined after the first laboratory tests. The de-focusing due to a temperature difference is 0.15 mm/degree, therefore we have to determine such working temperature with an accuracy between +0.7° and -2.0°. This temperature must not be lower than the temperature of the overall spectrograph to avoid dust deposition.